\documentclass[pra,twocolumn,floatfix,a4paper,superscriptaddress]{revtex4}
\usepackage{bm,color,graphicx,amsmath,txfonts}

\usepackage[colorlinks=true, allcolors=blue]{hyperref}
\begin{document}

\title{Sympathetic cooling of a radio-frequency LC circuit to its ground state \\ in an optoelectromechanical system}

%\author{xx}
\author{Nicola Malossi}
\affiliation{Physics Division, School of Science and Technology, University of Camerino, I-62032 Camerino (MC), Italy}
\affiliation{INFN, Sezione di Perugia, via A. Pascoli,
I-06123 Perugia, Italy}
\author{Paolo~Piergentili}
\affiliation{Physics Division, School of Science and Technology, University of Camerino, I-62032 Camerino (MC), Italy}
\affiliation{INFN, Sezione di Perugia, via A. Pascoli, I-06123 Perugia, Italy}
\author{Jie Li}
\affiliation{Kavli Institute of Nanoscience, Department of Quantum Nanoscience, Delft University of Technology, 2628CJ Delft, The Netherlands}
\affiliation{Zhejiang Province Key Laboratory of Quantum Technology and Device, Department of Physics and State Key Laboratory of Modern Optical Instrumentation, Zhejiang University, Hangzhou 310027, China}
%\author{Francesco Rasponi}
%\affiliation{Physics Division, School of Science and Technology, University of Camerino, I-62032 Camerino (MC), Italy}
\author{Enrico Serra}
\affiliation{Institute of Materials for Electronics and Magnetism, Nanoscience-Trento-FBK Division, I-38123 Povo, Trento, Italy}
\affiliation{INFN, Trento Institute for Fundamental Physics and Application, I-38123 Povo, Trento, Italy}
\affiliation{Dept. of Microelectronics and Computer Engineering /ECTM/DIMES, Delft University of Technology, Feldmanweg 17, 2628
CT Delft, The Netherlands}
\author{Riccardo~Natali}
\affiliation{Physics Division, School of Science and Technology, University of Camerino, I-62032 Camerino (MC), Italy}
\affiliation{INFN, Sezione di Perugia, via A. Pascoli, I-06123 Perugia, Italy}
\author{Giovanni~Di~Giuseppe}
\affiliation{Physics Division, School of Science and Technology, University of Camerino, I-62032 Camerino (MC), Italy}
\affiliation{INFN, Sezione di Perugia, via A. Pascoli, I-06123 Perugia, Italy}
\author{David Vitali}
\affiliation{Physics Division, School of Science and Technology, University of Camerino, I-62032 Camerino (MC), Italy}
\affiliation{INFN, Sezione di Perugia, via A. Pascoli, I-06123 Perugia, Italy}
\affiliation{CNR-INO, Largo Enrico Fermi 6, I-50125 Firenze, Italy}

\begin{abstract}

We present a complete theory for laser cooling of a macroscopic radio-frequency LC electrical circuit by means of an optoelectromechanical system, consisting of an optical cavity dispersively coupled to a nanomechanical oscillator, which is in turn capacitively coupled to the LC circuit of interest. The driven optical cavity cools the mechanical resonator which in turn sympathetically cools the LC circuit. We determine the optimal parameter regime where the LC resonator can be cooled down to its quantum ground state, which requires a large optomechanical cooperativity, and a larger electromechanical cooperativity. Moreover, comparable optomechanical and electromechanical coupling rates are preferable for reaching the quantum ground state.

\end{abstract}

\date{\today}
\maketitle

\section{Introduction}

Over the past decade, the experimental realization of quantum states of macroscopic objects has made significant progress in the fields of opto- and electromechanics. These include mechanical ground state cooling~\cite{OConnell2010,Teufel2011,Chan2011,Rossi2018,Qiu2020}, mechanical squeezing~\cite{Wollman2015,Pirkkalainen2015}, entanglement between mechanical, microwave and optical modes~\cite{Palomaki2013,Ockeloen2018,Riedinger2018,Barzanjeh2019,Chen2020}. Also facilitated by this progress, hybrid quantum systems \cite{Clerk2020} provide interesting opportunities and a variety of novel platforms for new technological applications. In particular, optoelectromechanical devices has received significant attention, especially in transducing radio-frequency (rf) and microwave signals to the optical domain~\cite{Taylor,Regal1,Barzanjeh,Bochmann,Andrews,PolzikNat,Balram,Vainsecher,Takeda,Midolo2018,Iman,Higginbotham,Simonsen,Forsch,Jiang,Fink,Han,Arnold,Chu}.

However, most of optoelectromechanical systems are using a GHz microwave resonator. Here, we focus onto the case of a MHz rf resonator, for which operation in the quantum domain is more difficult because, due to the lower resonance frequency, it is normally in a thermally excited state even at ultra-cryogenic temperatures. Radio-frequency signals in the MHz and kHz regimes are used in a large variety of research field and applications~\cite{Midolo2018}, ranging for example from astronomical signal detection at long wavelength (Astronomical Plasmas, sun activity and exoplanets research)~\cite{Bentum} to ultra-low Magnetic field Nuclear Magnetic resonance and imaging (SQUID coupled to an LC circuit)~\cite{Sarracanie}. Therefore the possibility of operating in a quantum regime at the MHz and even kHz range with extremely low noise can be advantageous either for positioning, timing and sensing (imaging) applications, and for more fundamental science applications, such as the sensitive detection of rf-signals of astrophysical nature.\\
Quantum operation of rf circuits requires cooling them close to their quantum ground state, and here we show that this can be achieved by appropriately engineering the interactions in a hybrid tripartite optoelectromechanical system. This result could be considered as a further example of quantum manifestation at macroscopic level, involving photons with macroscopic wavelengths and typically realised with macroscopically-sized circuit elements.

Laser cooling of an LC circuit via the intermediate coupling to a mechanical resonator has been first proposed in Ref.~\cite{Taylor}. Here we extend that analysis, showing that one can cool down the LC resonator to its quantum ground state, providing an alternative route to what has been recently demonstrated through the coupling to a superconducting qubit~\cite{Steele}, or to an ultra-cryogenic microwave cavity~\cite{Steele2}. In this paper we provide a detailed analysis of the system, by first determining its optimal working point, and then analysing its stationary state, focusing onto the parameter regime in which the rf LC resonant circuit can be ground state cooled. From a physical point of view, this occurs when the energy exchange process of the LC circuit with its own thermal reservoir is dominated by the exchange process with the much colder reservoir represented by the mechanical resonator cooled by the driven optical cavity mode. In more intuitive terms, the driven cavity cools the mechanical resonator which in turn sympathetically cools the rf resonator~\cite{sympa}. In general, we find that ground state cooling of the rf resonator is possible when the optomechanical cooperativity is large, and the electromechanical cooperativity is even larger. A preliminary study of the quantum behavior of the same optoelectromechanical system has been recently shown in Ref.~\cite{Jie}, which however focused only on the entanglement between the mechanical and the rf resonator. Ground state cooling and stationary entanglement are generally related quantum phenomena, but, as already verified in optomechanics~\cite{GenesMari,RMP}, they are optimised under quite different conditions.

The paper is organized as follows. In Sec.~\ref{model}, we introduce our tripartite optoelectromechanical system and provide its Hamiltonian and the corresponding Langevin equations. In Sec.~\ref{working} we determine the working point of the system and derive the linearized equations for the system quantum fluctuations. In Sec.~\ref{sec:steady} we show how to exactly solve these linearized equations and determine the steady state of the system, while in ~Sec.~\ref{approxtheory} we provide an approximate analytical theory for the steady state occupancy of the rf resonator. In ~Sec.~\ref{cooling} we describe the results and determine the optimal parameter regime for laser cooling the LC circuit to its quantum ground state. Then, in Sec.~\ref{sec:detection} we discuss in detail the challenges one has to face for an unambiguous detection of the stationary state of rf resonator, while Sec.~\ref{conc} is for concluding remarks.

\begin{figure}[t]
\hskip0cm\includegraphics[width=0.7\linewidth]{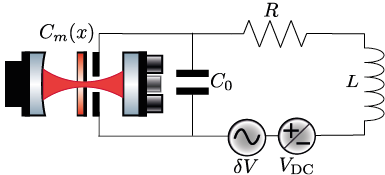}
\caption{Schematic description of the system. A metal coated nanomembrane is coupled via radiation pressure to a cavity field, and capacitively coupled to an rf resonant circuit via the position-dependent capacitance $C_m(x)$. The rf resonator is modelled as a lumped-element RLC series circuit with an additional tunable capacitance $C_0$ in parallel with $C_m(x)$, a resistance $R$, and an inductance $L$. The rf-circuit is driven by a DC bias $V_{\rm DC}$ and by the Johnson-Nyquist voltage noise $\delta V$.}
\label{fig1}
\end{figure}

\section{The system}
\label{model}

We consider a generic hybrid optoelectromechanical system, which consists of an optical cavity, a nanomechanical oscillator, and a radiofrequency (rf) resonant circuit. Different kind of systems and configurations have been already proposed and characterized experimentally~\cite{Taylor,Regal1,Barzanjeh,Bochmann,Andrews,PolzikNat,Balram,Vainsecher,Iman,Higginbotham,Forsch,Jiang,Fink,Han,Arnold} and the treatment presented here can be applied to all the cases in which the electromechanical coupling is capacitive, and the optomechanical coupling is dispersive. Nonetheless, in order to be more specific, we will refer to the configuration in which the optomechanical system is the membrane-in-the-middle (MIM) one~\cite{HarrisNat,HarrisNJP,Kimble,Biancofiore,Karuza}, i.e., a driven optical Fabry-Per\'ot cavity with a thin semitransparent membrane inside. The membrane is metalized~\cite{Taylor,Andrews,PolzikNat,Iman,Regal2} and capacitively coupled via an electrode to an LC resonant circuit formed by a coil and additional capacitors, see Fig.~\ref{fig1}.
The Hamiltonian of the system can be written in general as the sum of an optical, mechanical and electrical term,
\begin{equation}\label{hin}
  \hat{H}=\hat{H}_{\rm opt}+\hat{H}_{\rm mech}+\hat{H}_{\rm LC},
\end{equation}
where
\begin{eqnarray}
   \hat{H}_{\rm opt} &=& \hbar \omega(x)\:\hat{a}^\dag \hat{a} + i\hbar E\left(\hat{a}^\dag e^{-i\omega_L t}-\hat{a} e^{i\omega_L t}\right), \\
   \hat{H}_{\rm mech} &=& \frac{\hat{p}^2}{2 m}+\frac{m \omega_{0}^{2} \hat{x}^2}{2}, \\
   \hat{H}_{\rm LC} &=& \frac{\hat{\phi}^2}{2 L}+\frac{\hat{q}^2}{2 C(\hat{x})}-\hat{q} \hat{V}.
\end{eqnarray}
In the optical contribution we consider a specific cavity mode, with photon annihilation (creation) operator $\hat{a}$ ($\hat{a}^\dag$), with the usual bosonic commutation relations $[\hat{a},\hat{a}^\dag]=1$, which is driven by a laser of frequency $\omega_L$ and input power $\mathcal{P}$. Consequently, the driving rate can be written as $E = \sqrt{2\kappa_{\rm in} \mathcal{P}/\hbar\omega_L}$, with $\kappa_{in}$ the cavity amplitude decay rate through the input port. The mechanical Hamiltonian corresponds to a resonator with mass $m$, displacement operator $\hat{x}$ and conjugated momentum $\hat{p}$, with commutation relation $[\hat{x},\hat{p}]=i\hbar$, which is associated to a given vibrational mode of the metalized membrane with bare frequency $\omega_0$. The dispersive optomechanical coupling arises due to the dependence of the cavity mode frequency $\omega(\hat{x})$ upon the membrane displacement $\hat{x}$, as discussed in Refs.~\cite{HarrisNat,HarrisNJP,Kimble,Biancofiore,Karuza}.

The electrical contribution $\hat{H}_{\rm LC}$ refers to the rf resonator, which we will describe here as a lumped-element series RLC circuit (see Fig.~1), whose dynamical variables are given by the concatenated flux $\hat{\phi}$ and the total capacitor charge $\hat{q}$, with the canonical commutation relation $[\hat{q},\hat{\phi}]=i\hbar$. We have also included a driving term associated with the possibility to control the circuit via a voltage bias $\hat{V}$. The electromechanical coupling is capacitive and it arises from the displacement dependence of the effective circuit capacitance $C(\hat{x})$. In the case of the chosen optoelectromechanical setup based on a metalized membrane, such as those of Refs.~\cite{Taylor,Andrews,PolzikNat,Iman}, one can write
\begin{equation}\label{capac}
  C(\hat{x})=C_0+C_m(\hat{x}) = C_0 + \frac{\varepsilon_0 A_{\rm eff}}{h_0+\hat{x}},
\end{equation}
i.e., the effective capacitance is the parallel of a tunable capacitance $C_0$ with the capacitor formed by the metalized membrane together with the electrodes in front of it. As shown in Refs.~\cite{PolzikNat,Iman}, we can assume a parallel plate model and define the effective area $A_{\rm eff}$ of the membrane capacitor; $h_0$ is the steady state distance between the membrane and the electrodes, in the absence of any bias voltage and cavity laser driving, while $\varepsilon_0 $ is the vacuum permittivity.

A realistic RLC circuit is always quite involved, with its behaviour determined by a number of parasitic capacitances and resistances whose values depend upon the specific circuit implementation. However, the simplified description adopted here in terms of the three lumped-element effective quantities, the inductance $L$, the resistance $R$, and the capacitance $C(\hat{x})$, is possible and perfectly suited for our purposes. In fact, our goal is to laser cool the rf circuit via its quasi-resonant interaction with the mechanical resonator, and the dynamical behaviour is essentially determined by the frequency components around the rf resonance peak, which is characterised by two easily measurable quantities, the rf-resonant frequency $\omega_{LC}^{(0)}$ and the width, $\gamma_{LC}$, of the uncoupled resonator. The two quantities define the rf-circuit quality factor $Q_{LC}=\omega_{LC}^{(0)}/\gamma_{LC}$, which must be large enough, $Q_{LC}\gg 1$, in order to achieve an appreciable cooling~\cite{Taylor}. A third circuit quantity that can be directly measured is its effective inductance $L$, which can be obtained from the low frequency behaviour of the circuit. Therefore, since in a high-\textit{Q} series RLC circuit one has $\omega_{LC}^{(0)}=1/\sqrt{LC}$ and $\gamma_{LC}=R/L$, once that the value of the circuit inductance $L$ has been measured, one can define the other two effective circuit parameters as
\begin{eqnarray}\label{lumpeff}
  C(0) &=& C_0+\frac{\varepsilon_0 A_{\rm eff}}{h_0}\equiv\frac{1}{L \left[\omega_{LC}^{(0)}\right]^2}, \\
  R &\equiv & L \gamma_{LC}.
\end{eqnarray}
The full quantum dynamics of the system and its stationary state can be determined from the Heisenberg-Langevin equations of the system which are obtained from the Hamiltonian of Eq.~(\ref{hin}) and by including fluctuation-dissipation processes for the three resonators, which in the frame rotating at the laser frequency $\omega_L$, are given by
\begin{eqnarray}
  \dot{\hat{x}}&=&\hat{p}/m, \label{eqHL1}\\
  \dot{\hat{p}} &=& -m \omega_0^2 \hat{x} - \gamma_m \hat{p} - \hbar\,\frac{\partial \omega}{\partial \hat{x}} (\hat{x})\hat{a}^\dag \hat{a} \nonumber \\
	&&- \frac{\hat{q}^2}{2} \frac{\varepsilon_0 A_{\rm eff}}{\left[C_0(h_0+\hat{x})+\varepsilon_0 A_{\rm eff}\right]^2} + \hat{F} , \\
  \dot{\hat{q}} &=& \frac{\hat{\phi}}{L},\\
  \dot{\hat{\phi}} &=& -\frac{\hat{q}}{C_0+\varepsilon_0 A_{\rm eff}/(h_0+\hat{x})}-\gamma_{LC} \hat{\phi}+V_{\rm DC}+\delta \hat{V},\\
  \dot{\hat{a}} &=& i[\omega_L-\omega(\hat{x})]\,\hat{a} - \kappa \hat{a}
   			+\! E \!+\! \sqrt{2 \kappa_{\rm in}} \hat{a}_{\rm in}\!+\!\sqrt{2 \kappa_{\rm ex}} \hat{a}_{\rm ex}, \label{eqHL5}
\end{eqnarray}
where $\gamma_m$ is the mechanical damping rate, and $\kappa = \kappa_{\rm in}+\kappa_{\rm ex}$ is the total cavity amplitude decay rate, given by the sum of the decay rate though the input port $\kappa_{\rm in}$ and the decay rate through all the other ports $\kappa_{\rm ex}$. The latter optical loss processes are associated with the corresponding input noise operators $\hat{a}_{\rm in}$ and $\hat{a}_{\rm ex}$, which are uncorrelated and whose only nonzero correlation is $\langle \hat{a}_{\rm j}(t) \, \hat{a}_{\rm j}^{\dag}(t')\rangle = \delta(t-t')$, ${\rm j = in,ex}$.

We have included two zero-mean noise terms in the equations: $\hat{F}(t)$ is the Langevin force operator which accounts for the Brownian motion of the mechanical oscillator, whose symmetrized correlation function is in general equal to \cite{Gardiner,Markov}
\begin{eqnarray}\label{browncorre}
 && \frac{1}{2}\langle \hat{F}(t)\hat{F}(t')+\hat{F}(t')\hat{F}(t) \rangle \\
 && \qquad\quad=m\gamma_m\int \frac{d\omega}{2\pi}\cos \omega (t-t')\hbar \omega \coth\left(\frac{\hbar \omega}{2 k_B T}\right), \nonumber
\end{eqnarray}
which, in the case of a large mechanical quality factor $Q_m= \omega_0/\gamma_m \,\, {\gg}\, 1$ valid here, can be approximated with the Markovian expression~\cite{Markov},
$\langle \hat{F}(t)\hat{F}(t')+\hat{F}(t')\hat{F}(t) \rangle/2  \simeq m \gamma_m \hbar \omega_0(2 \bar{n}_m+1) \delta(t-t')$, where $\bar{n}_m=[e^{\hbar \omega_0/k_B T}-1]^{-1}$ is the equilibrium mean thermal phonon number, with $k_B$ the Boltzmann constant and $T$ the environmental temperature.
We have also rewritten the external bias voltage as $\hat{V}(t) = V_{\rm DC}+\delta \hat{V}(t)$, i.e., the sum of a DC bias and the Johnson-Nyquist voltage noise operator $\delta \hat{V}$ with autocorrelation function~\cite{Gardiner,JN},
\begin{eqnarray}\label{brownnyquist}
  &&\frac{1}{2}\langle \delta \hat{V}(t)\delta \hat{V}(t')+\delta \hat{V}(t') \delta \hat{V}(t) \rangle \\
 && \qquad\quad=R\int \frac{d\omega}{2\pi}\cos \omega (t-t')\hbar \omega \coth\left(\frac{\hbar \omega}{2 k_B T}\right), \nonumber
\end{eqnarray}
which again, in the case of a large LC quality factor can be approximated with the Markovian expression
$\langle \delta \hat{V}(t)\delta \hat{V}(t')+\delta \hat{V}(t') \delta \hat{V}(t) \rangle/2  \simeq  R \hbar \omega_{LC}^{(0)}( 2\bar{n}_{LC}+1) \delta(t-t')$, where $\bar{n}_{LC}=[e^{\hbar \omega_{LC}^{(0)}/k_B T_{LC}}-1]^{-1}$ is the mean thermal rf photon number. We have assumed in general $T_{LC} \neq T$ because the rf circuit tends to pick up ambient noise and the effective rf noise temperature can be larger than ambient temperature.

\section{Working point and linearized dynamics of the quantum fluctuations}
\label{working}

In order to look for the possibility to reach the quantum regime for the macroscopic rf resonator, we have to evaluate the stationary quantum fluctuations around the classical steady state of the system, which is obtained by replacing all the operators in the Heisenberg-Langevin equations Eqs.~(\ref{eqHL1})-(\ref{eqHL5}) with the corresponding average values, neglecting all noise terms, and setting all the derivatives to zero. In this way one defines the working point of the system, which is determined by the two external drivings, i.e., the laser driving rate $E$ and the DC bias voltage $V_{\rm DC}$. If stability conditions are satisfied (see Appendix), the steady state is characterized by the cavity mode in a coherent state with amplitude $\alpha_s$, the membrane in an equilibrium position displaced by $x_s$, the rf circuit with no current, and the capacitor with a stationary charge $q_s$. Using the fact that $p_s=\phi_s=0$, one can express the working point parameters in terms of $x_s$ only, i.e.,
\begin{eqnarray}
\alpha_s &=& \frac{E}{\kappa+ i \Delta}  ,\label{Ave} \\
q_s &=& C(x_s)V_{\rm DC},\label{Ave1}
\end{eqnarray}
where $\Delta = \omega(x_s)-\omega_L$ is the effective cavity mode detuning, and it is the parameter which is actually fixed in an experiment by the cavity stabilisation system. The static membrane displacement $x_s$ is the solution of the equilibrium condition for the three forces applied to the membrane, i.e., the membrane elastic force,  the electrostatic force and the radiation pressure force ,
\begin{equation}\label{eqxs}
  m \omega_0^2 x_s = -\frac{\varepsilon_0 A_{\rm eff}V_{\rm DC}^2}{2(h_0+x_s)^2}-\hbar\frac{\partial \omega}{\partial x} (x_s)n_{\rm cav},
\end{equation}
where $n_{\rm cav}=|\alpha_s|^2 =E^2/(\kappa^2+\Delta^2)$ is the intracavity photon number.

The quantum fluctuations dynamics can be described with very good approximation by linearizing the exact Heisenberg-Langevin equations Eqs.~(\ref{eqHL1})-(\ref{eqHL5}) around the classical stationary state defining the system working point, i.e., by keeping only first order terms in such fluctuations. In fact, if stability conditions are satisfied, the system dynamics will not significantly depart from the steady state defined above, and higher order terms in the fluctuation operators can be neglected~\cite{GenesMari,RMP}. It is convenient to express these equations in terms of dimensionless fluctuation operators, scaled by the corresponding quantum zero-point fluctuation units, i.e., by redefining
\begin{eqnarray}\label{rescale1}
  \hat{x} &\to & x_s + x_{\rm zpf} \delta \hat{x} = x_s + \sqrt{\frac{\hbar}{m\omega_0}} \delta \hat{x}, \\
  \hat{p} &\to & p_s + p_{\rm zpf} \delta \hat{p} = \sqrt{\hbar m\omega_0} \delta \hat{p}, \\
  \hat{q} &\to & q_s + q_{\rm zpf} \delta \hat{q} = q_s + \sqrt{\frac{\hbar}{L\omega_{LC}^{(0)}}} \delta \hat{q}, \\
  \hat{\phi} &\to & \phi_s + \phi_{\rm zpf} \delta \hat{\phi} = \sqrt{\hbar L\omega_{LC}^{(0)}} \delta \hat{\phi} , \label{rescale4}
\end{eqnarray}
so that the commutation relations are rewritten as $[\delta \hat{x},\delta \hat{p}]=[\delta \hat{q},\delta \hat{\phi}]=i$. By introducing also the two intracavity quadrature fluctuation operators
\begin{eqnarray}\label{quadcav}
  \delta \hat{X} &=& \frac{\delta \hat{a} e^{i\theta}+\delta \hat{a}^\dag e^{-i\theta}}{\sqrt{2}},  \\
  \delta \hat{Y} &=& \frac{\delta \hat{a} e^{i\theta}-\delta \hat{a}^\dag e^{-i\theta}}{i\sqrt{2}} ,
\end{eqnarray}
where $\theta = \arctan\Delta/\kappa$, one gets the following linearized Heisenberg-Langevin equations
\begin{eqnarray}\label{QLE2a}
\delta \dot{\hat{x}}&=& \omega_0 \delta \hat{p},   \\
\delta \dot{\hat{p}}&=& - \frac{\omega_m^2}{\omega_0} \delta \hat{x} - \gamma_m \delta \hat{p} + G \delta \hat{X} - g \delta \hat{q}  + \hat{\xi},   \\
\delta \dot{\hat{q}}&=& \omega_{LC}^{(0)} \delta \hat{\phi},   \\
\delta \dot{\hat{\phi}}&=& -  \frac{\omega_{LC}^2}{\omega_{LC}^{(0)}} \delta \hat{q} - \gamma_{LC} \delta \hat{\phi} - g \delta \hat{x}  + \delta \hat{{\cal V}},   \\
\delta \dot{\hat{X}}&=& \Delta \delta \hat{Y} - \kappa \delta \hat{X} + \sqrt{2 \kappa} \hat{X}_{\rm vac} ,  \\
\delta \dot{\hat{Y}}&=& - \Delta \delta \hat{X} - \kappa \delta \hat{Y} + G \delta \hat{x} + \sqrt{2 \kappa} \hat{Y}_{\rm vac} . \label{QLE2z}
\end{eqnarray}
We have introduced the two relevant coupling rates, the optomechanical coupling rate
\begin{equation}\label{optocoup}
  G = -x_{\rm zpf}\frac{\partial \omega(x_s)}{\partial x} \sqrt{2n_{\rm cav}},
\end{equation}
and the electromechanical coupling rate
\begin{equation}\label{electrocoup}
  g = \frac{\varepsilon_0 A_{\rm eff}V_{\rm DC}}{C(x_s)(h_0+x_s)^2\sqrt{mL\omega_{LC}^{(0)}\omega_0}}.
\end{equation}
\begin{figure}[t]
\hskip0cm\includegraphics[width=0.95\linewidth]{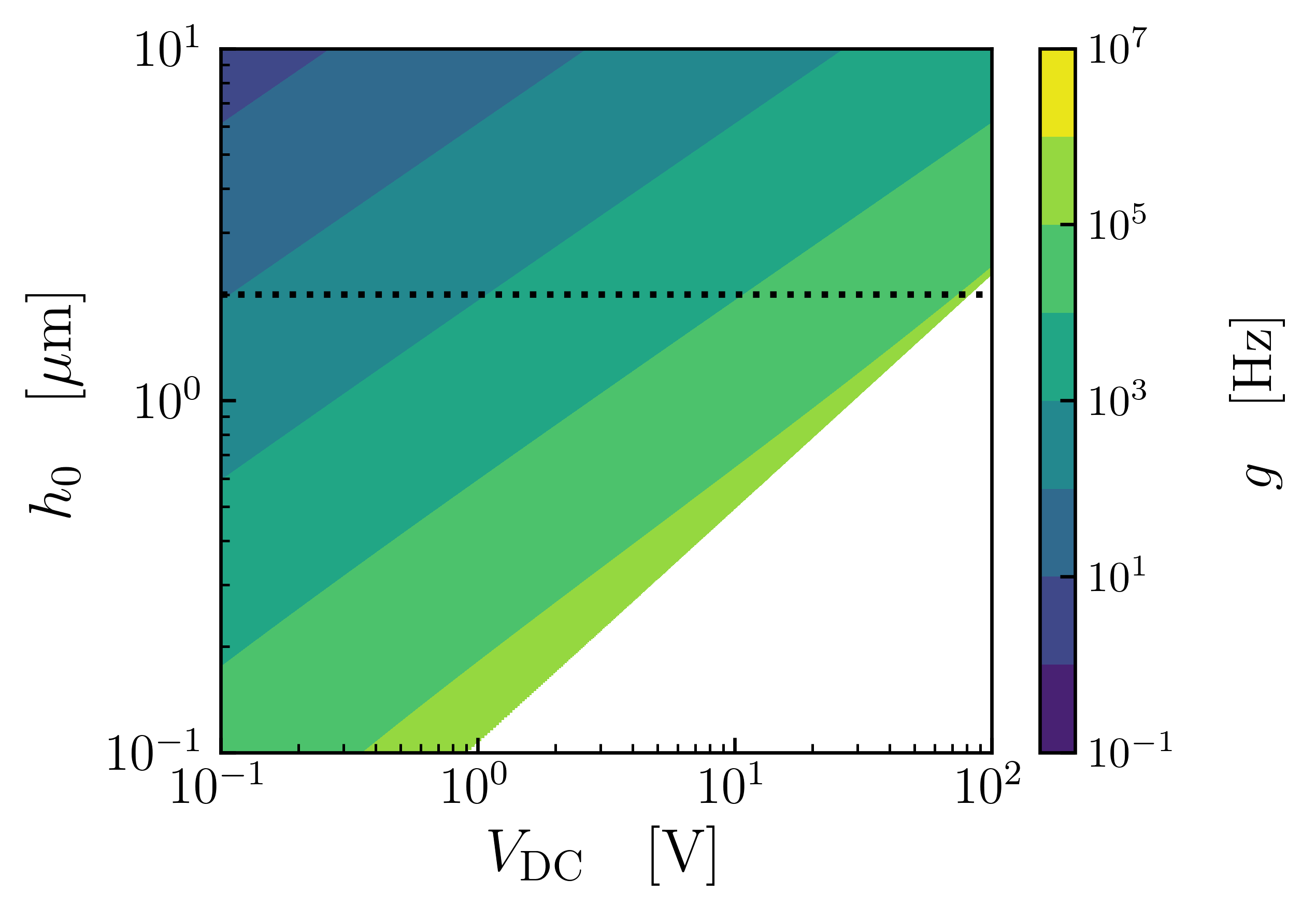}
\caption{Electro-mechanical coupling $g$ versus the DC voltage $V_{\rm DC}$ and the membrane-electrode distance $h_{0}$. The black dotted line indicates the value of $h_{0}$ which is used in the plots of Sec.~\protect\ref{cooling}, corresponding to $2$ $\mathrm{\mu m}$. The other electro-mechanical parameters are: $\omega_0/2\pi= \omega_{LC}^{(0)}/2\pi=1$ MHz,
%for the mechanical resonator frequency,
$Q_{m}=10^{6}$, $m=0.7 \times 10^{-10}$ kg, $L=1$ mH,
%for the effective inductance of the LC circuit,
$A_{\rm eff}=1.1 \times 10^{-7}$ m$^2$.}
%for the effective area of the membrane capacitor.}
\label{fig2}
\end{figure}
We notice that both the mechanical and the LC resonance frequencies, $\omega_0$ and $\omega_{LC}^{(0)}$ respectively, are modified when the cavity is driven and the DC voltage bias is applied, acquiring new values: from Eq.~(\ref{capac}) and Eq.~(\ref{lumpeff}) one has
\begin{equation}\label{lumpeff2}
 \omega_{LC}^2= \left[L C(x_s)\right]^{-1}=\left[L \left(C_0+\frac{\varepsilon_0 A_{\rm eff}}{h_0+x_s}\right)\right]^{-1},
\end{equation}
while the modified mechanical resonance frequency $\omega_m$ is given by the expression
\begin{equation}\label{omegaren}
  \omega_m^2= \omega_0^2+\frac{\hbar}{m}\frac{\partial^2 \omega(x_s)}{\partial x^2} n_{\rm cav}-\frac{V_{\rm DC}^2\varepsilon_0 A_{\rm eff}}{m(h_0+x_s)^3}.
\end{equation}
We recall that the system is stable provided that $\omega_m^2 >0$ and the latter expression shows that there is a maximum value for $V_{\rm DC}$, the pull-in voltage, beyond which the effective mechanical frequency $\omega_m$ becomes imaginary and the membrane is pulled onto the other electrode of the capacitor (see Appendix A). We also notice that for physically interesting parameter regimes, the shift $x_s$ may be not negligible with respect to $h_0$ and tends to $-h_0/3$ when approaching the pull-in voltage (see Appendix A). As a consequence, due to Eq.~(\ref{eqxs}) and Eq.~(\ref{electrocoup}), the coupling $g$ has a nonlinear dependence upon $V_{\rm DC}$, and it never surpasses a maximum value when $V_{\rm DC}$ approaches its maximum value $V_{\rm pull}$. This is explicitly shown in Fig.~\ref{fig2}, where the electromechanical coupling $g$ is shown versus the electrode distance $h_0$ and $V_{\rm DC}$.
The stationary membrane shift $x_s$ is determined by the equilibrium between the mechanical stress, the electrostatic force and the radiation force. As shown in Appendix B, where we provide the explicit expressions for the membrane-in-the-middle case based on the treatment of Ref. \cite{Biancofiore}, the contribution of the radiation force on $x_s$ is negligible.

Finally we have also introduced rescaled noise operators: i) the mechanical thermal noise term $\hat{\xi}(t)=\hat{F}(t)/p_{\rm zpf}$, with symmetrized autocorrelation function (in the high $Q_m$ Markovian limit)
\begin{equation}\label{rescabrown}
\frac{1}{2}\left\langle \hat{\xi}(t)\hat{\xi}(t')+\hat{\xi}(t')\hat{\xi}(t) \right\rangle  = \gamma_m (2 \bar{n}_m+1) \delta(t-t');
\end{equation}
ii) the rescaled Nyquist noise operator on the rf circuit $\delta \hat{{\cal V}}=\delta \hat{V}(t)/\phi_{\rm zpf}$, with symmetrized autocorrelation function (in the high $Q_{LC}$ Markovian limit)
\begin{equation}\label{rescanyqui}
 \frac{1}{2} \langle \delta \hat{{\cal V}}(t)\delta \hat{{\cal V}}(t')+\delta \hat{{\cal V}}(t')\delta \hat{{\cal V}}(t) \rangle  =\gamma_{LC} (2 \bar{n}_{LC}+1) \delta(t-t');
\end{equation}
iii) the two vacuum optical noises
\begin{eqnarray}
  \hat{X}_{\rm vac} &=& \frac{1}{\sqrt{2\kappa}}\left[\sqrt{\kappa_{\rm in}}\left(\hat{a}_{\rm in}e^{i\theta}+\hat{a}_{\rm in}^{\dag}e^{-i\theta}\right)\right.\\
  &&\qquad\quad\left.+\sqrt{\kappa_{\rm ex}}\left(\hat{a}_{\rm ex}e^{i\theta}+\hat{a}_{\rm ex}^{\dag}e^{-i\theta}\right)\right], \nonumber \\
  \hat{Y}_{\rm vac} &=& \frac{-i}{\sqrt{2\kappa}}\left[\sqrt{\kappa_{\rm in}}\left(\hat{a}_{\rm in}e^{i\theta}-\hat{a}_{\rm in}^{\dag}e^{-i\theta}\right)\right.\\
  &&\qquad\quad\left. +\sqrt{\kappa_{\rm ex}}\left(\hat{a}_{\rm ex}e^{i\theta}-\hat{a}_{\rm ex}^{\dag}e^{-i\theta}\right)\right], \nonumber
\end{eqnarray}
which are uncorrelated and possess the same autocorrelation function
\begin{eqnarray}\label{vacuumcorre}
&& \frac{1}{2}\left\langle \hat{X}_{\rm vac}(t)\hat{X}_{\rm vac}(t')+\hat{X}_{\rm vac}(t')\hat{X}_{\rm vac}(t) \right\rangle  \\
&&\qquad\quad= \frac{1}{2}\left \langle \hat{Y}_{\rm vac}(t)\hat{Y}_{\rm vac}(t')+\hat{Y}_{\rm vac}(t')\hat{Y}_{\rm vac}(t) \right\rangle = \frac{1}{2}\delta(t-t'). \nonumber
\end{eqnarray}

\section{Determination of the steady state}
\label{sec:steady}

The linearized Heisenberg-Langevin Equations in~(\ref{QLE2a})-(\ref{QLE2z}) can be rewritten in the following compact matrix form
\begin{equation}\label{compaeq}
\dot{\hat{u}}(t)=A \hat{u}(t)+\hat{n}(t),
\end{equation}
where $\hat{u}(t)=[\delta \hat{x}(t), \delta \hat{p}(t),\delta \hat{q}(t),\delta \hat{\phi}(t),\delta \hat{X}(t), \delta \hat{Y}(t)]^T$ is the column vector of fluctuations (the superscript $T$ denotes transposition), $\hat{n}(t) =[0, \hat{\xi}(t),0,\delta \hat{{\cal V}}(t),\sqrt{2\kappa}\hat{X}_{\rm vac}(t),\sqrt{2\kappa}\hat{Y}_{\rm vac}(t)]^T$ is the corresponding column vector of noises, and $A$ is the matrix
\begin{equation}\label{dynmat}
  A=\left(\begin{array}{cccccc}
    0 & \omega_0 & 0 & 0 & 0 & 0\\
     -\frac{\omega_m^2}{\omega_0} & -\gamma_m & -g & 0 & G & 0 \\
    0 & 0 & 0 & \omega_{LC}^{(0)} & 0 & 0\\
     -g & 0 & - \frac{\omega_{LC}^2}{\omega_{LC}^{(0)}} & -\gamma_{LC} & 0 & 0 \\
    0 & 0 & 0 & 0 &-\kappa & \Delta \\
    G & 0 & 0 & 0 & -\Delta & -\kappa
  \end{array}\right).
\end{equation}
The formal solution of Eq.~(\ref{compaeq}) is
$$ \hat{u}(t)=M(t) \hat{u}(0)+\int_0^t ds M(s) \hat{n}(t-s),$$
where $M(t)=\exp\{A t\}$. The system is stable and
reaches its steady state for $t \to \infty$ when all the eigenvalues of $A$ have negative real parts so that $M(\infty)=0$.
Here we will consider the parameter region where the driven cavity mode cools the mechanical resonator, corresponding to a driving laser red-detuned with respect to the cavity, $\Delta > 0$~\cite{RMP}. Within this parameter region, the stability condition is violated only at very large values of the optomechanical coupling $G$, which are detrimental for cooling, correspond to the onset of optical bistability~\cite{Dorsel}, and which are of no interest here.

In the linearized regime, the steady state of the tripartite optoelectromechanical system can be fully characterized because the noise terms are zero-mean quantum Gaussian noises, and as a consequence, the steady state of the system is a zero-mean tripartite Gaussian state, fully determined by its $6 \times 6 $ correlation
matrix (CM) $ V_{ij}=\left(\langle \hat{u}_i(\infty)\hat{u}_j(\infty)+ \hat{u}_j(\infty)\hat{u}_i(\infty)\rangle\right)/2$.

Starting from Eq.~(\ref{compaeq}), this
steady state CM can be determined in two equivalent ways, either as an integral (see Eqs.~(\ref{Vfin})-(\ref{cm3}) below), or as solution of a matrix equation (see Eq.~(\ref{lyap}) below). Using the Fourier transforms $\hat{u}_i(\omega)$ of $\hat{u}_i(t)$, one has
\begin{equation}
V_{ij}(t)\!=\!\!\!\!\int\!\!\!\! \int\!\!\!\frac{d\omega d\omega'}{(2\pi)^2}e^{-it(\omega+\omega')}\!\frac{1}{2}\!\left\langle \!\hat{u}_i(\omega) \hat{u}_j(\omega')\!+\!
\hat{u}_j(\omega')\hat{u}_i(\omega)\!\right\rangle. \label{defV}
\end{equation}
Then, by Fourier transforming Eq.~(\ref{compaeq}) and the correlation functions of the noises in the Markovian limit, Eqs.~(\ref{rescabrown}), (\ref{rescanyqui}) and (\ref{vacuumcorre}), one gets
\begin{equation}\label{lemma}
\!\!\!\frac{\left\langle \hat{u}_i(\omega) \hat{u}_j(\omega')\!+\! \hat{u}_j(\omega')\hat{u}_i(\omega)\right\rangle}{2}\!\!
=\!\!\left[M(\omega)D M(\omega')^{T}\right]_{ij}\!\!\delta(\omega\!+\!\omega'),
\end{equation}
where we have defined the $6 \times 6 $ matrix
\begin{equation} M(\omega)=\left(i\omega+A\right)^{-1}, \label{invea}
\end{equation}
and the diagonal diffusion matrix
\begin{equation}\label{diffuom}
  D=\left(\begin{array}{cccccc}
    0 & 0 & 0 & 0 & 0 & 0 \\
     0 & \gamma_m (2 \bar{n}_m+1) & 0 & 0 & 0 & 0 \\
     0 & 0 & 0 & 0 & 0 & 0\\
     0 & 0 & 0 & \gamma_{LC} (2 \bar{n}_{LC}+1) & 0 & 0 \\
     0 & 0 & 0 & 0 & \kappa & 0 \\
     0 & 0 & 0 & 0 & 0 & \kappa
  \end{array}\right).
\end{equation}
The $\delta(\omega+\omega')$ factor is a consequence of the stationarity of the noises, and
inserting Eq.~(\ref{lemma}) into Eq.~(\ref{defV}), one gets the following expression for the stationary correlation matrix
\begin{equation}
V^{\infty}=\int_{-\infty}^{\infty} \frac{d\omega}{2\pi} M(\omega)D M(\omega)^{\dagger}, \label{Vfin}
\end{equation}
which can be equivalently rewritten as an integral in the time domain as
\begin{equation} \label{cm3}
V^{\infty} =\int_0^{\infty} dt  M(t)D M(t)^{T}.
\end{equation}
From the latter expression one can derive an alternative way to get the stationary CM $V^{\infty}$. In fact, when the stability conditions are satisfied [$M(\infty)=0$],
one can verify, by an explicit integration, that Eq.~(\ref{cm3}) is equivalent to the following Lyapunov equation for the steady-state CM,
\begin{equation} \label{lyap}
AV^{\infty}+V^{\infty}A^{T}=-D.
\end{equation}
This is a linear equation for $V^{\infty}$ which can be analytically solved, but the general exact expression of the matrix elements is too cumbersome and will not be
reported here. We have adopted this latter method, and the numerical analysis and the plots of Sec.~\ref{cooling} are obtained from the numerical solution of Eq.~(\ref{lyap}).

In this paper we are interested only in the stationary state of the rf resonator and in its stationary energy in particular, which is equal to
\begin{eqnarray}
U_{LC} &=&  \frac{\hbar \omega_{LC}^{(0)}}{2} \left[\langle \delta \hat{q}^2 \rangle + \langle \delta \hat{\phi}^2 \rangle \right]=\frac{\hbar \omega_{LC}^{(0)}}{2} \left(V^{\infty}_{33}+V^{\infty}_{44} \right)  \label{ulc} \\
 & \equiv & \hbar \omega_{LC}^{(0)} \,\left(\bar{n}_{LC}^{\rm eff}+\frac{1}{2}\right),\nonumber
\end{eqnarray}
where $\bar{n}_{LC}^{\rm eff}$ is the effective mean occupation number of the LC oscillator.

\section{Approximate expression for the rf resonator occupancy}\label{approxtheory}

One can adapt standard resolved sideband cooling theory of optomechanical systems for the derivation of an approximate expression of the stationary occupancy of the rf resonator.
In the case of a standard optomechanical system, the stationary occupancy of the mechanical resonator, far from the strong coupling regime, can be very well approximated as \cite{TJK,FM,Genes}:
\begin{equation}\label{nmeffmech}
  \bar{n}_m^{\rm eff}=\frac{\gamma_m \bar{n}_m +\Gamma_m\bar{n}_c+ A_+}{\gamma_m + \Gamma_m},
\end{equation}
where $\bar{n}_c\simeq0$ is the mean excitation of the optical reservoir at zero temperature, $\Gamma_m =A_{-}-A_{+}>0$ is the net laser cooling rate, with
\begin{equation}\label{scattering}
A_{\pm }=\frac{G^{2}\kappa /2}{\kappa ^{2}+(\Delta \pm \omega _{m})^{2}}
\end{equation}
the scattering rates into the Stokes ($A_{+}$) and anti-Stokes ($A_{-}$) sidebands, corresponding respectively to the absorption or emission of a mechanical vibrational quantum. Eq.~(\ref{nmeffmech}) can be seen as the result of the balance between the two energy exchange processes involving the mechanical resonator: i) the one with rate $\gamma_m$ with its thermal reservoir with $\bar{n}_m$ mean excitations; ii) the other one with rate $\Gamma_m$ with the effective optical reservoir at zero temperature ($\bar{n}_c\simeq 0$) represented by the driven and decaying cavity, and which is responsible for cooling. The scattering rate $A_+$ is responsible for the quantum back-action limit associated with the quantum fluctuations of the radiation pressure force.

In the optoelectromechanical system under study, the rf resonator we are interested in is directly coupled to the mechanical resonator, which is in turn coupled to the driven optical cavity. In the proposal of Ref.~\cite{Taylor} one can laser cool the rf resonator by driving on the red sideband of the optical cavity as in the usual optomechanical sideband cooling, and then exploiting the resonant electromechanical interaction in order to extend cooling to the rf circuit. This is why one can view this process as sympathetic cooling \cite{sympa} of the LC resonator by means of the laser cooling of the mechanical resonator.

An equivalent description of the desired cooling process is the following: the rf resonator is cooled because the energy exchange process at rate $\gamma_{LC}$ with its own thermal reservoir with $\bar{n}_{LC}$ mean excitations is dominated by the exchange process with the \emph{much colder ``polariton'' reservoir} represented by the mechanical resonator hybridized with the driven optical cavity mode in the regime of efficient sideband cooling. This latter effective reservoir is characterised by an effective decay rate $\gamma_m^{\rm eff}=\gamma_m+\Gamma_m$, a nonzero mean number of excitations $ \bar{n}_m^{\rm eff}$ [see Eq.~(\ref{nmeffmech})], and the LC resonator will scatter polaritons into the corresponding Stokes and anti-Stokes sidebands with rates that are respectively given by
\begin{equation}\label{scattering2}
A_{\pm }^{LC}=\frac{g^{2}\gamma_m^{\rm eff }}{\left(\gamma_m^{\rm eff }\right)^2+4\left(\omega_{m} \pm \omega_{LC}\right)^{2}}.
\end{equation}
An intuitive explanation of the present expression is the fact that, when comparing with the optomechanical case of Eq.~(\ref{scattering}), the rate $\gamma_m^{\rm eff}/2$ plays the role of the cavity decay rate $\kappa$, and the electromechanical coupling $g$ plays the role of the optomechanical coupling $G$. As a consequence one has an effective polariton cooling rate
\begin{equation}\label{netpolariton}
\Gamma_{LC} =A_{-}^{LC}-A_{+}^{LC}>0.
\end{equation}
One can then apply the same arguments used for deriving Eq.~(\ref{nmeffmech}) to the present situation, and arrive at the following expression for the rf resonator occupancy
\begin{equation}\label{nmeffLC}
  \bar{n}_{LC}^{\rm eff}=\frac{\gamma_{LC} \bar{n}_{LC} + \Gamma_{LC}\bar{n}_{m}^{\rm eff}+ A_+^{LC}}{\gamma_{LC} + \Gamma_{LC}}.
\end{equation}
This is the desired approximation we were looking for. It works in the optimal regime for sideband cooling, that is, $\Delta \sim \omega_m \sim \omega_{LC} > \kappa > G$ as well as $ \gamma_m^{\rm eff}/2 \sim G^2/4\kappa > g$. From Eq.~(\ref{nmeffLC}) one can see that the rf resonator cannot be cooled more than the mechanical resonator and that therefore at best one can achieve $ \bar{n}_{LC}^{\rm eff} \sim \bar{n}_{m}^{\rm eff}$. The latter condition is achieved when $\Gamma_{LC} \sim A_-^{LC} \gg A_+^{LC},\gamma_{LC}$, which is obtained at resonance $\Delta \sim \omega_m \sim \omega_{LC} \gg \gamma_m^{\rm eff } \sim G^2/2\kappa$, when $2g^2 \kappa/G^2 \gg \gamma_{LC}$. Defining the two relevant cooperativities, the optomechanical cooperativity ${\cal C}_{om} = G^2/2\kappa\gamma_m$ and the electromechanical cooperativity ${\cal C}_{em} = g^2/\gamma_{LC}\gamma_m$, the necessary condition to achieve simultaneous ground state cooling,  $ \bar{n}_{LC}^{\rm eff} \sim \bar{n}_{m}^{\rm eff} < 1$, can be written as
\begin{equation}\label{cooperativity}
  {{\cal C}_{em} \gg \cal C}_{om} \gg 1.
\end{equation}
This latter condition for the cooperatives can be satisfied only for an LC circuit with a large enough value of its quality factor, so that $\gamma_{LC} \ll g,\kappa$, because the electromechanical coupling $g$ cannot be too large with respect to $G^2/\kappa$ for the validity of the above expressions. Nonetheless, the results of Sec.~\ref{cooling} based on the exact numerical solution of the Lyapunov equation of Eq.~(\ref{lyap}) show that cooling of the rf resonator close to the quantum regime is possible also when the above assumptions are not fully satisfied and Eq.~(\ref{nmeffLC}) is not too accurate.

\section{Results for the cooling of the LC resonator}
\label{cooling}

\begin{figure}[t]
\hskip0cm\includegraphics[width=0.99\linewidth]{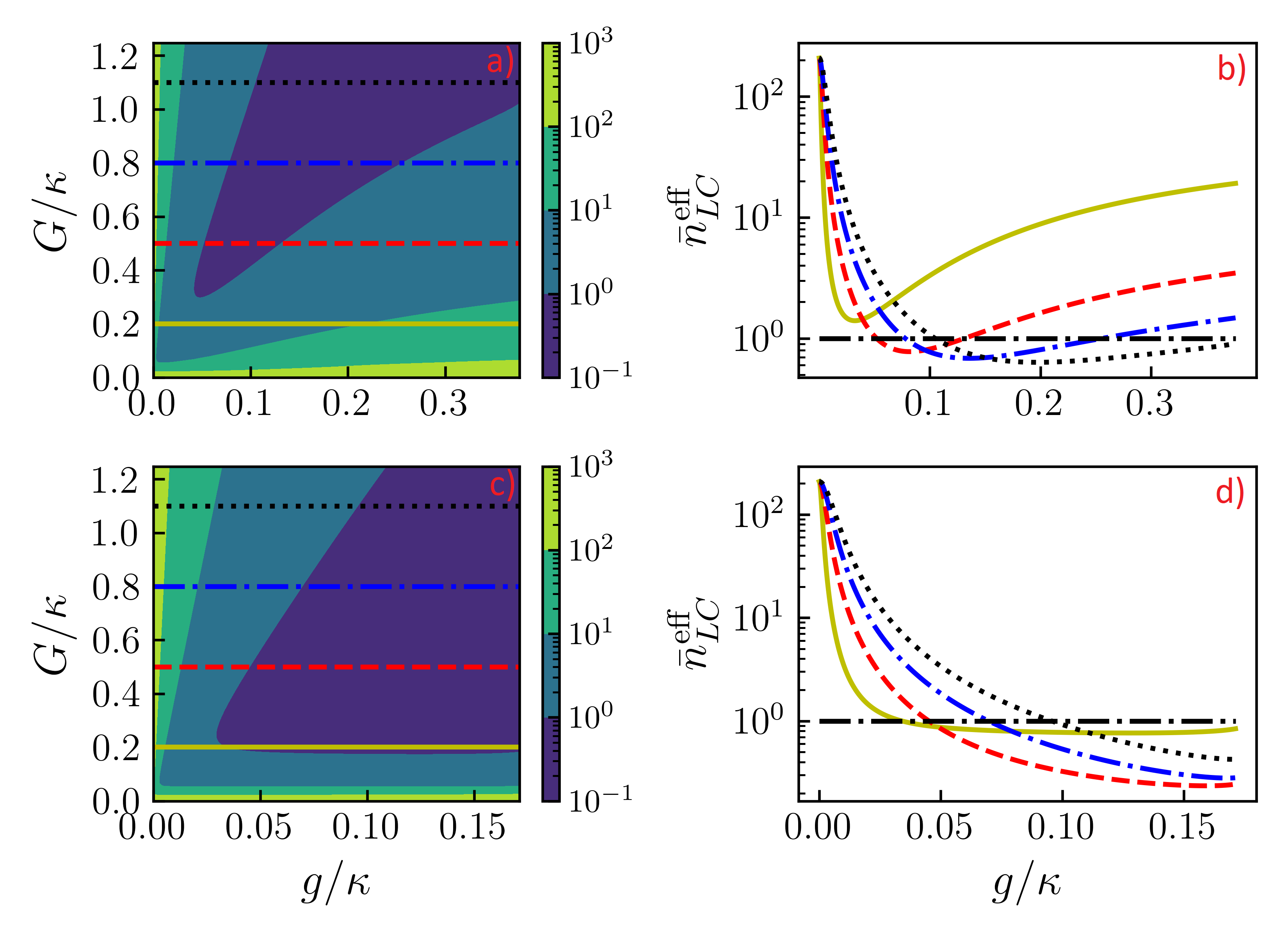}
%\hskip0cm\includegraphics[width=0.99\linewidth]{figWtenutoinrisonanza.png}
\caption{\textbf{a)} and \textbf{c)}: Stationary LC circuit occupancy $\bar{n}_{LC}^{\rm eff}$ from the solution of the Lyapunov equation, Eq.~(\ref{lyap}), as a function of the scaled electromechanical coupling $g/\kappa$, and of the scaled optomechanical coupling $G/\kappa$. \textbf{b)} and \textbf{d)}: $\bar{n}_{LC}^{\rm eff}$ versus $g/\kappa$, at fixed values of the optomechanical coupling rate, corresponding to the horizontal lines in a)-c): $G/\kappa=0.2$ (full yellow lines), $G/\kappa=0.5$ (dashed red lines), $G/\kappa=0.8$ (dash-dotted blue lines), $G/\kappa=1.1$ (dotted black lines). The upper plots a) and b) refer to the resonance of the bare mechanical and LC frequencies, $\omega_0 = \omega_{LC}^{(0)}$, while the lower plots c) and d) refer to the resonance of the \emph{effective} mechanical and LC frequencies, $\omega_m = \omega_{LC}$. We have chosen a quality factor of the rf resonator $Q_{LC}=4 \times 10^4$ and a temperature $T=10$ mK (see text for the other system parameters). The dash-dotted black horizontal line in b) and d) refers to $\bar{n}_{LC}^{\rm eff} =1$.}
\label{fig3}
\end{figure}

\begin{figure}[t]
\hskip0cm\includegraphics[width=0.95\linewidth]{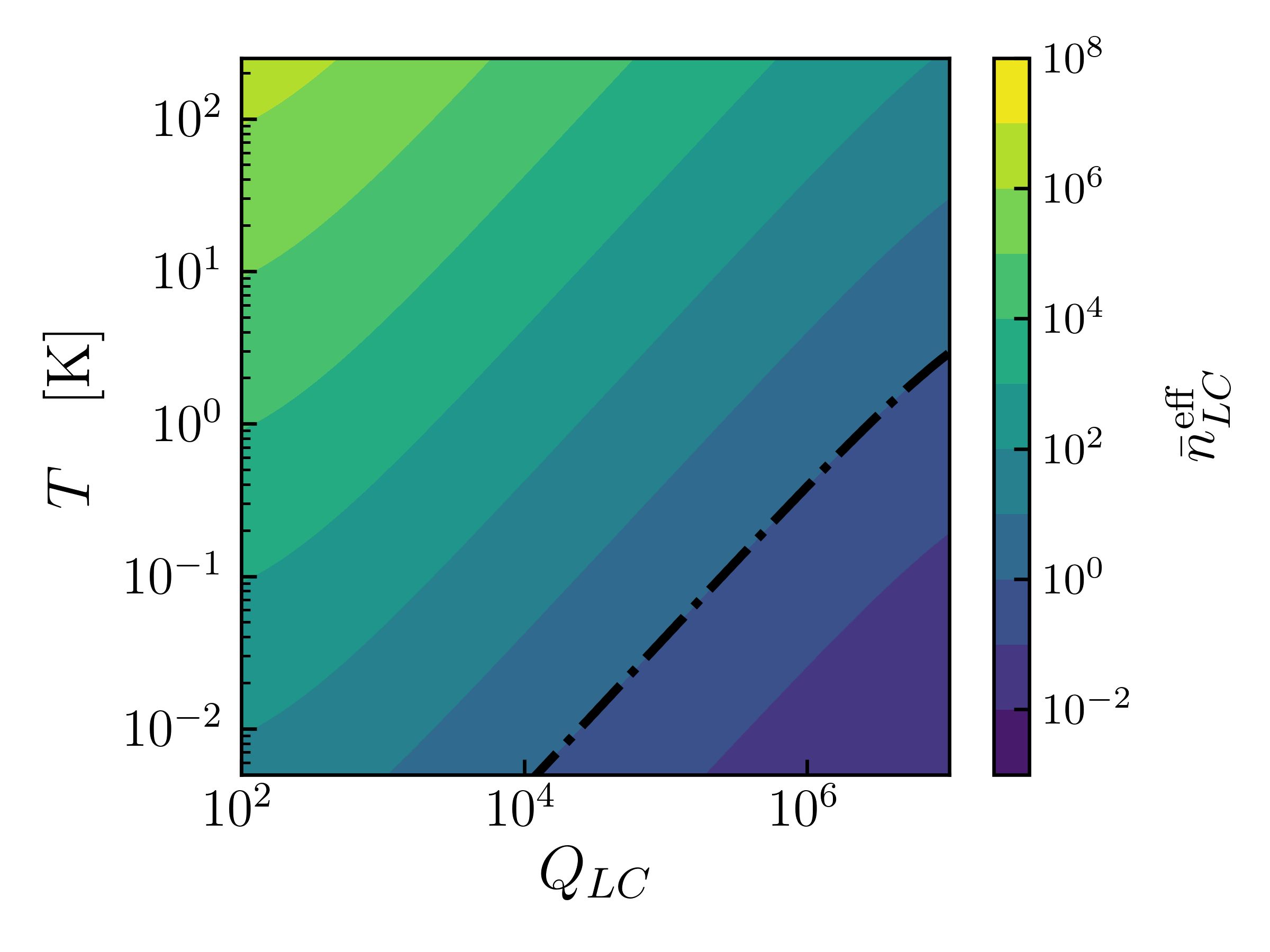}
\caption{Stationary LC circuit occupancy $\bar{n}_{LC}^{\rm eff}$ from the solution of the Lyapunov equation, Eq.~(\ref{lyap}), as a function of the rf resonator quality factor $Q_{LC}$ and of temperature $T$ (we have assumed here $T=T_{LC}$), for a chosen value of the optomechanical coupling, $G/\kappa=0.8$ (dash-dotted blue line of Fig.~\ref{fig3}), choosing $\omega_0 = \omega_{LC}^{(0)}$, and by fixing, for any given pair of values of $Q_{LC}$ and $T$, the optimal value of the electromechanical coupling $g$ minimising $\bar{n}_{LC}^{\rm eff}$. The dash-dotted line refers to the upper bound of the ``quantum region'', $\bar{n}_{LC}^{\rm eff}=1$. }
\label{fig4}
\end{figure}

\begin{figure}[t]
\hskip0cm\includegraphics[width=0.85\linewidth]{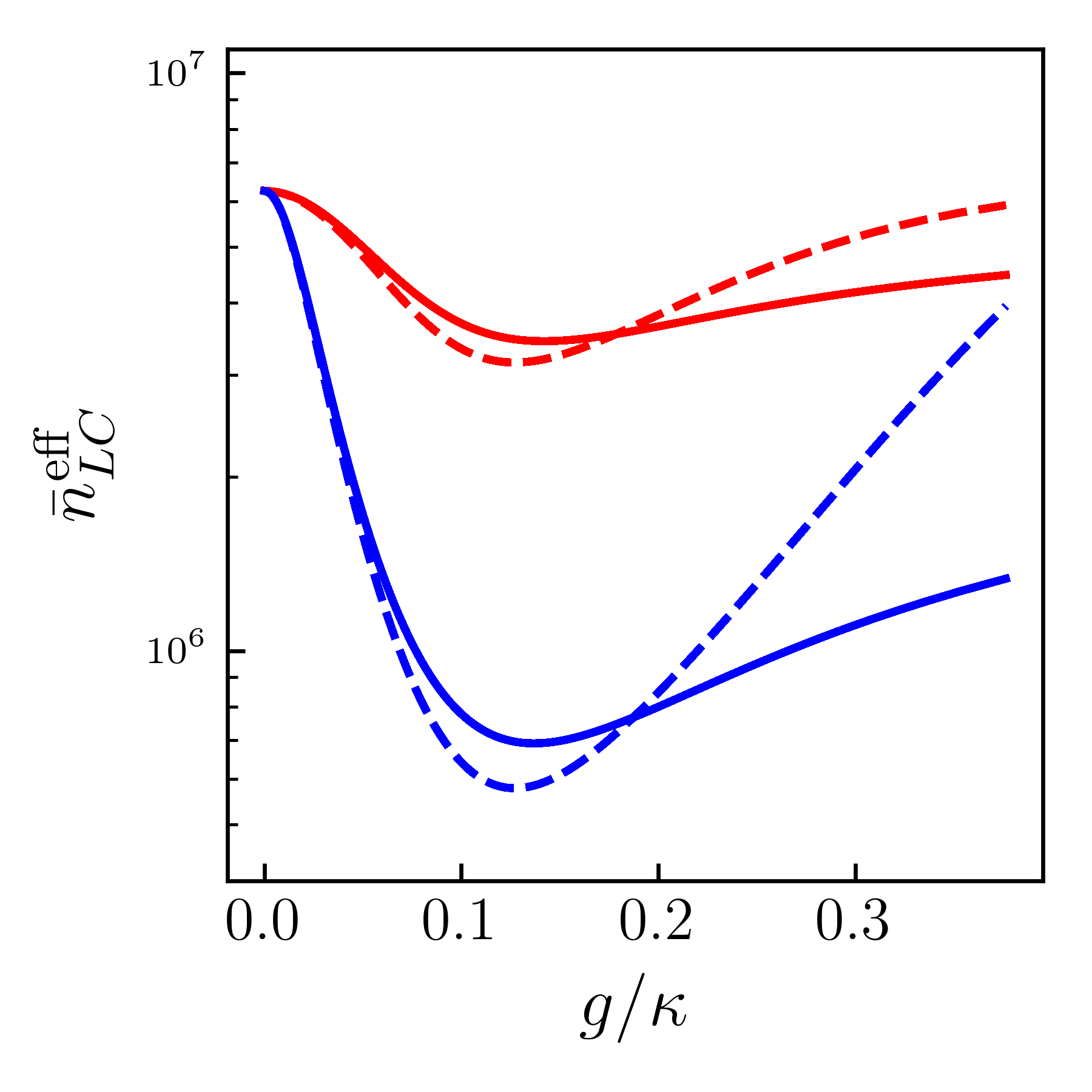}
\caption{LC resonator photon occupation number $\bar{n}_{LC}^{\rm eff}$ versus $g/\kappa$, at the same value of the optomechanical coupling, $G/\kappa=0.8$, chosen in Fig.~\ref{fig4}, at temperature $T=T_{LC}=300$K, choosing $\omega_0 = \omega_{LC}^{(0)}$, and for two different values of the quality factor, $Q_{LC}=10^2$ (red full and dashed upper curves), and $Q_{LC}=10^3$ (blue full and dashed lower curves).
Full lines refer to the exact numerical solution of Eq.~(\ref{lyap}), while dashed lines refer to the approximate treatment of Sec.~\ref{approxtheory} [see Eq.~(\ref{nmeffLC})].}
\label{fig5}
\end{figure}

Let us now determine the optimal parameter conditions under which one can cool a MHz rf circuit down to its quantum ground state. We show the main results in Fig.~\ref{fig3} and Fig.~\ref{fig4}, where we apply the exact treatment of Eqs.~(\ref{lyap})-(\ref{ulc}). Then in Fig.~\ref{fig5} we compare these results with the approximate treatment of Sec.~\ref{approxtheory} and the corresponding analytical prediction of Eq.~(\ref{nmeffLC}), showing a satisfactory agreement between them.

As we have seen above, the most relevant parameters one has to optimise are the optomechanical coupling $G$, the electromechanical coupling $g$, the ambient temperatures $T$ and $T_{LC}$ (which here will be taken to be identical for simplicity), and the quality factor of the rf resonant circuit, $Q_{LC}$.

The other parameters will be kept fixed and corresponding to typical experimental values for a metalized membrane-in-the-middle configuration \cite{PolzikNat,Iman}, that is,
laser optical wavelength $\lambda=1064 $ nm, membrane effective mass $m=0.7 \times 10^{-10}$ kg, membrane intensity reflectivity (in the non-metalized section) $R=0.4$, bare mechanical resonance frequency $\omega_0= 2\pi \times 1 $ MHz, mechanical quality factor $Q_{m}=10^{6}$, optical cavity length $L_c=8 \times 10^{-3}$ m, optical cavity finesse $F=5 \times 10^{4}$, yielding a total optical cavity amplitude decay rate $\kappa=2\pi \times 374.74$ kHz. We have also chosen $\kappa_{\rm in}=0.4 \kappa $ and a laser driving around the red mechanical sideband, that is, $\Delta \simeq \omega_0$. Finally, we have chosen an equivalent circuit inductance $L=1$ mH, and a membrane capacitor with an effective area $A_{\rm eff}=1.1 \times 10^{-7}$  m$^2$ and distance between the electrodes equal to $h_{0}=2$ $\mu$m. As a consequence, the two coupling rates $G$ and $g$, and the corresponding system working point, can be tuned by varying the two external parameters, the driving laser power ${\cal P}$ which fixes the intracavity photon number $n_{\rm cav}$, and the DC voltage bias $V_{\rm DC}$.

In Fig.~\ref{fig3} we assume a cryogenic temperature $T=T_{LC}=10$ mK, and a rf resonator quality factor $Q_{LC}=4 \times 10^4$. Then in Fig.~\ref{fig3}a)-\ref{fig3}b) we take identical bare frequencies of the LC and mechanical resonators, $\omega_{LC}^{(0)}=\omega_{0}$, and display the behaviour of the stationary rf circuit occupancy $\bar{n}_{LC}^{\rm eff}$ predicted by Eqs.~(\ref{lyap})-(\ref{ulc}). In Fig.~\ref{fig3}a) $\bar{n}_{LC}^{\rm eff}$ is plotted versus the scaled electromechanical and optomechanical coupling rates, $g/\kappa$ and $G/\kappa$ respectively, while in Fig.~\ref{fig3}b) we plot $\bar{n}_{LC}^{\rm eff}$ versus $g/\kappa$ at four fixed values of $G/\kappa$ corresponding to the horizontal lines with the same color and style in Fig.~\ref{fig3}a). These plots clearly show that an experimentally achievable parameter region exists where it is possible to reach the quantum regime with an LC resonator occupation number below 1.

However, the analysis of Sec.~\ref{approxtheory} predicts that optimal laser cooling of the LC resonator occurs when the \emph{modified} mechanical and rf resonance frequencies of Eqs.~(\ref{lumpeff2})-(\ref{omegaren}) are resonant, $\omega_m = \omega_{LC}$, rather than their bare counterparts, $\omega_{LC}^{(0)}$ and $\omega_{0}$.
Therefore, in Fig.~\ref{fig3}c)-\ref{fig3}d) we display the same plots of Fig.~\ref{fig3}a)-\ref{fig3}b) but now under the optimal resonance condition $\omega_m = \omega_{LC}$, which can always be obtained by adjusting, at each working point, the value of $\omega_{LC}$ by means of the tuning capacitance $C_0$ [see Eq.~(\ref{lumpeff2})]. As expected, cooling of the LC resonator is improved, with a wider parameter region where laser cooling is efficient, even though the qualitative behavior is not appreciably modified.

By comparing the upper and lower plots in Fig.~\ref{fig3}, we notice that the coupling $g$ spans a shorter interval of possible values in the latter resonant case. In fact, for a given choice of $h_{0}$ and $A_{\rm eff}$, $g$ is already upper limited by the maximum $V_{\rm DC}$ that can be applied before the pull-in effect (see Fig.~\ref{fig2}); when we impose the resonance condition $\omega_m = \omega_{LC}$, since the mechanical frequency $\omega_m$ decreases due to the applied DC bias [see Eq.~(\ref{omegaren})], the overall capacitance of the LC circuit must be increased, implying a further limitation to the value of $g$ [see Eq.~(\ref{electrocoup})].
From Fig.~\ref{fig3} we can conclude that ground state cooling of the LC circuit is achievable for comparable values of the coupling rates $G$ and $g$, with the latter ones obtained with a voltage bias $V_{\rm DC}$ always far enough from the pull-in voltage (see Fig.~\ref{fig2}).

The more demanding experimental condition assumed in Fig.~\ref{fig3} is the one on the rf resonator quality factor, because typical values are in the range $10^2 \lesssim Q_{LC} \lesssim 10^3$~\cite{PolzikNat,Steele}, even though very recently Ref.~\cite{Steele2} demonstrated a rf resonator with $Q_{LC} \sim 1.7 \times 10^4$. For this reason in Fig.~\ref{fig4} we have studied the LC circuit occupancy $\bar{n}_{LC}^{\rm eff}$ also as a function of the rf resonator quality factor $Q_{LC}$ and of temperature $T=T_{LC}$, at a fixed value of the optomechanical coupling, $G/\kappa=0.8$ (blue dash-dotted line of Fig.~\ref{fig3}) and $\omega_{LC}^{(0)}=\omega_{0}$ for simplicity. At each point of the plot, we have chosen the optimal value of the electromechanical coupling $g$ minimising $\bar{n}_{LC}^{\rm eff}$. As expected, cooling improves for lower temperature and larger $Q_{LC}$, and the dash-dotted line sets the boundary of the ``quantum region'' where $\bar{n}_{LC}^{\rm eff}\leq 1$.

Finally, in Fig.~\ref{fig5} we compare the exact numerical result of Eq.~(\ref{lyap}) with the approximate analytical theory developed in Sec.~\ref{approxtheory}. We show the mean rf photon number $\bar{n}_{LC}^{\rm eff}$ versus $g/\kappa$ for the numerical solution (full lines) and for the approximate analytical theory of Eq.~(\ref{nmeffLC}) (dashed lines) at $T=300$ K, choosing again $\omega_{LC}^{(0)}=\omega_{0}$, and for two different, realistic values of the LC quality factor, $Q_{LC}=10^2$ (red upper curves), and $Q_{LC}=10^3$ (blue lower curves).
The approximate theory well reproduces the numerical solution for relatively low values of the electromechanical coupling $g$, up to the value roughly corresponding to the minimum occupancy. For larger $g$, the prediction of Eq.~(\ref{nmeffLC}) increases more than the numerical solution, which is somehow expected because the approximated theory is valid as long as $g$ is not larger than the effective optomechanical decay rate $G^2/2\kappa$. Nonetheless, the approximate theory provides a very good estimate of the achievable cooling limit as well as of the $g$-interval where the minimum rf-photon occupancy could be achieved.

\section{Detection of the rf resonator steady state}
\label{sec:detection}

The effective mean photon number of the rf circuit at the steady state can be measured following two ways: i) measuring directly the rf voltage signal between two points of the circuit; ii) measuring the optical output of the cavity and trying to get information about the rf circuit state from it. In both cases these measurements are carried out in the frequency domain and therefore here we will focus on the solution of the Fourier transform of the Heisenberg-Langevin equations, Eq.~(\ref{compaeq}). This solution has been already given in compact form in Sec.~\ref{sec:steady}, but it will be convenient to re-express it in more physical terms using effective susceptibilities.

Solving separately the two quadrature equations for each mode in equations from Eq.~(\ref{QLE2a}) to Eq.~(\ref{QLE2z}), we get
\begin{eqnarray}
&&\chi_c^{-1}(\omega)\delta \hat{X} (\omega)\!\!=\!\!G \delta \hat{x}(\omega)+ \!\! \sqrt{2 \kappa}\left[\frac{\kappa -i \omega}{\Delta}\!\hat{X}_{\rm vac}(\omega)\! +\!\hat{Y}_{\rm vac} (\omega) \right]\!,\quad\\
&&\chi_m^{-1}(\omega) \delta \hat{x} (\omega)\!=\! G \delta \hat{X} (\omega) - g \delta \hat{q} (\omega)  + \hat{\xi}(\omega)\label{QLE3}, \\
&&\chi_{LC}^{-1}(\omega) \delta \hat{q} (\omega)\!=\! - g \delta \hat{x} (\omega)  + \delta \hat{{\cal V}} (\omega),\\
\nonumber
\end{eqnarray}
where $\chi_c(\omega)$, $\chi_m(\omega)$, and $\chi_{LC}(\omega)$ are the natural susceptibilities of the cavity, mechanical, and electrical modes, respectively, given by
\begin{equation}\label{3chi}
\begin{split}
\chi_c(\omega)&= \frac{\Delta}{\Delta^2 + (\kappa - i \omega)^2 }    , \\
\chi_m(\omega)&= \frac{\omega_0}{\omega_m^2 - \omega^2 - i \gamma_m \omega }    , \\
\chi_{LC}(\omega)&= \frac{\omega_{LC}^{(0)}}{\omega_{LC}^2 - \omega^2 - i \gamma_{LC} \omega }    . \\
\end{split}
\end{equation}
The mutual interactions among the three modes lead to the modification of their natural susceptibilities. Inserting $\delta \hat{X} (\omega)$ and $\delta \hat{q} (\omega)$ in Eq.~\eqref{QLE3} into the equation for $\delta \hat{x} (\omega)$, we obtain
\begin{eqnarray}
\left[\chi_m^{\rm eff}(\omega)\right]^{-1} \delta \hat{x} (\omega) &=&  \chi_c(\omega)  G \! \sqrt{2 \kappa} \left[\frac{\kappa -i \omega}{\Delta} \hat{X}_{\rm vac} (\omega) + \!\! \hat{Y}_{\rm vac} (\omega) \right] \nonumber  \\
&+& \hat{\xi} (\omega) - \chi_{LC}(\omega) g  \delta \hat{{\cal V}} (\omega), \label{eqx}
\end{eqnarray}
where $\chi_m^{\rm eff}(\omega)$ is the effective mechanical susceptibility, defined by
\begin{equation}
\left[\chi_m^{\rm eff}(\omega)\right]^{-1} =  \chi_{mc}^{-1}(\omega) - g^2  \chi_{LC}(\omega) ,
\end{equation}
with
\begin{equation}
\chi_{mc}^{-1}(\omega) =  \chi_{m}^{-1}(\omega) - G^2 \chi_c(\omega),
\end{equation}
where $\chi_{mc}(\omega)$ is the effective mechanical susceptibility in the presence of {\it only} the optomechanical interaction. Eq.~\eqref{eqx} together with $\delta \hat{p} (\omega)= -i(\omega/\omega_0) \delta \hat{x} (\omega)$ provides the mechanical response of the system to external perturbations.

Following the same approach, for the electrical mode we obtain
\begin{eqnarray}\label{eqq}
\left[\chi_{LC}^{\rm eff}(\omega)\right]^{-1}\!\! &&\!\!\!\!\delta \hat{q} (\omega)=  \delta \hat{{\cal V}} (\omega)  - \chi_{mc}(\omega)g\\
&&\!\!\!\times  \bigg\{  \chi_c(\omega)  G \! \sqrt{2 \kappa} \, \left[\frac{\kappa -i \omega}{\Delta} \hat{X}_{\rm vac} (\omega) +  \hat{Y}_{\rm vac} (\omega) \right]  + \hat{\xi} (\omega) \bigg\},\nonumber
\end{eqnarray}
where $\chi_{LC}^{\rm eff}(\omega)$ is the effective rf circuit susceptibility, given by
\begin{equation}\label{chiLCeff}
[\chi_{LC}^{\rm eff}(\omega)]^{-1} = \chi_{LC}^{-1}(\omega) - g^2 \chi_{mc}(\omega).
\end{equation}
In the same way Eq.~\eqref{eqq} together with $\delta \hat{\phi} (\omega)= -i(\omega/\omega_{LC}) \delta \hat{q} (\omega)$ provides the rf response of the system to external perturbations.

Eq.~(\ref{ulc}) can be rewritten as
\begin{equation}\label{ulc2}
  \bar{n}_{LC}^{\rm eff} = \frac{\langle \delta \hat{q}^2 \rangle + \langle \delta \hat{\phi}^2 \rangle -1 }{2},
\end{equation}
that is, the effective stationary rf photon number can be expressed in terms of the dimensionless charge and flux variances. In turn, using Eq.~(\ref{Vfin}), these variances are given by the integral over the corresponding noise spectra
\begin{eqnarray}
\label{deltaq}
  \!\!\langle \delta \hat{q}^2 \rangle\!\!\! &=&\!\!\!\!\! \int_{-\infty}^{+\infty}\!\!\! \frac{d \omega}{2\pi} \left[M(\omega)D M(\omega)^{\dagger}\right]_{33}\!\!\!=\!\!\int_{-\infty}^{+\infty}\!\!\! \frac{d \omega}{2\pi} S_{\!\delta q}(\omega), \\
  \!\!\langle \delta \phi^2 \rangle\!\!\! &=&\!\!\!\!\! \int_{-\infty}^{+\infty}\!\!\! \frac{d \omega}{2\pi} \left[M(\omega)D M(\omega)^{\dagger}\right]_{44}\!\!\!=\!\!\int_{-\infty}^{+\infty}\!\!\! \frac{d \omega}{2\pi} \frac{\omega^2}{\omega_{LC}^2} S_{\!\delta q}(\omega). \label{deltaphi}
\end{eqnarray}
Therefore laser cooling of the rf resonator can be experimentally verified by measuring the charge noise spectrum $ S_{\delta q}(\omega)$, which can be explicitly written in terms of the effective susceptibilities defined above as
\begin{equation}
S_{\!\delta q} (\omega) = \big| \chi_{LC}^{\rm eff}(\omega) \big|^2 \, \bigg[  g^2 \big| \chi_{mc}(\omega) \big|^2 \, \Big( S_{rp} + S_{\!\xi}  \Big)  +  S_{\!\delta {\cal V}}  \bigg], \label{spectrasusc}
\end{equation}
where $S_{\!rp}$ is the radiation pressure noise spectral contribution
\begin{equation}\label{rpspe}
S_{\!rp} (\omega) = G^2 \kappa  \frac{\Delta^2 + \kappa^2 + \omega^2}{ (\Delta^2 + \kappa^2 - \omega^2)^2 + 4 \kappa^2 \omega^2  }  ,
\end{equation}
and $S_{\!\xi}$ and $S_{\!\delta {\cal V}}$ are, respectively, the Brownian force noise and the voltage noise spectra, which are constant, white, contributions due to the Markovian approximation made on the Brownian and Johnson-Nyquist noise,
\begin{eqnarray}
S_{\!\xi} &=&  \gamma_m (2 \bar{n}_m+1),   \label{essexi}\\
S_{\!\delta {\cal V}}  &=&   \gamma_{LC} (2 \bar{n}_{LC}+ 1) . \label{essedeltav}
\end{eqnarray}
The charge noise spectrum can be measured by measuring either the voltage noise across the circuit capacitance $C(x_s)$, $\delta V_C(\omega)$, or the voltage noise across the circuit inductance $L$, $\delta V_L(\omega)$. Since $\delta V_C(\omega)=[q_{\rm zpf}/C(x_s)]\delta q(\omega)$, $\delta V_L(\omega)=-(\omega^2 \phi_{\rm zpf}/\omega_{LC})\delta q(\omega)$, and both measurements will be affected by an imprecision noise originating from the environment and the detection apparatus, one can write for the two cases
\begin{eqnarray}
  S_{\!\delta V_C} &=& \frac{q_{\rm zpf}^2}{C(x_s)^2}S_{\delta q}(\omega)+S_{imp}^C, \label{speccvolt}\\
  S_{\!\delta V_L} &=& \frac{\omega^4 \phi_{\rm zpf}^2}{\omega_{LC}^2}S_{\delta q}(\omega)+S_{imp}^L, \label{speclvolt} \\
  & = & \frac{\omega^4}{\omega_{LC}^4}\frac{q_{\rm zpf}^2}{C(x_s)^2}S_{\delta q}(\omega)+S_{imp}^L, \nonumber
\end{eqnarray}
where $S_{imp}^C$ and $S_{imp}^L$ are the voltage imprecision noise spectra for the measurement cases, we have assumed that they are uncorrelated from the other noise terms, and we have used Eq.~(\ref{lumpeff}) and Eq.~(\ref{rescale4}) in Eq.~(\ref{speclvolt}). These expressions show that measured voltage noise spectra can provide the stationary photon occupancy of the rf resonant circuit provided that the spectra are properly calibrated and, above all, that the imprecision noise is small enough so not to alter the evaluation of the area below the measured spectrum [see Eqs.~(\ref{ulc2})-(\ref{deltaphi})].

We can express this condition on the imprecision noise $S_{imp}^{C,L}$ in more quantitative terms by exploiting the fact that the shape of the charge spectrum $S_{\delta q}(\omega)$ is determined by the effective LC susceptibility of Eq.~(\ref{chiLCeff}). The parameter regime which is optimal for cooling the LC resonator is distinct from the strong electromechanical regime where the electrical and mechanical modes hybridize yielding two spectrally resolved peaks. On the contrary, in the cooling regime of interest here, the effective LC susceptibility $\chi_{LC}^{\rm eff}$ is characterized by a single peak, significantly broadened by the interaction with the optomechanical system, and therefore one can approximate $\chi_{LC}^{\rm eff}$ as a standard susceptibility with modified effective frequency $\omega_{LC}^{\rm eff}$ and damping $\gamma_{LC}^{\rm eff}$ \cite{Genes,Dantan},
\begin{equation}\label{chiLCeff2}
\left|\chi_{LC}^{\rm eff}(\omega)\right|^2 \simeq  \frac{\left[\omega_{LC}^{(0)}\right]^2}{\left[\left(\omega_{LC}^{\rm eff}\right)^2 - \omega^2\right]^2 + \left(\omega\ \gamma_{LC}^{\rm eff}\right)^2 }  ,
\end{equation}
where
\begin{equation}
\omega_{LC}^{\rm eff} \simeq  \sqrt{ \omega_{LC}^2 + \frac{g^2 \kappa^2}{G^2} } \simeq \omega_{LC},
\end{equation}
under typical experimental conditions, and
\begin{equation}
\gamma_{LC}^{\rm eff} \simeq   \gamma_{LC} + \Gamma_{LC},
\end{equation}
in agreement with the analysis of Sec.~\ref{approxtheory}. Therefore the charge noise spectrum $S_{\delta q}(\omega)$ is peaked at $\omega \simeq \omega_{LC}^{\rm eff} \simeq \omega_{LC}$, and, using Eqs.~(\ref{spectrasusc})-(\ref{essedeltav}), one can write its maximum value with very good approximation as
\begin{eqnarray}
S_{\delta q}^{\rm peak}
\simeq&
\hspace{-2.75cm}
S_{\delta q}(\omega_{LC}) = \frac{1}{ (\gamma_{LC}^{\rm eff})^2}\left\{\gamma_{LC}(2 \bar{n}_{LC}+1) \right. \\
&\quad\left.+\frac{g^2 \omega_0^2}{(\omega_m^2-\omega_{LC}^2)^2+(\omega_{LC}\, \gamma_{m}^{\rm eff})^2}\left[\gamma_m(2\bar{n}_m+1)+\frac{G^2 (2\omega_{LC}^2+\kappa^2)}{ \kappa (4\omega_{LC}^2+\kappa^2)}\right]\right\},\nonumber
\label{spectrapeak}
\end{eqnarray}
where we have approximated also the effective optomechanical susceptibility in the Lorentzian-like form \cite{Genes,Dantan}
\begin{equation}\label{chimceff}
\left|\chi_{mc}(\omega)\right|^2 \simeq  \frac{\omega_{0}^2}{\left(\omega_{m}^2 - \omega^2\right)^2 +  \left(\omega\,\gamma_{m}^{\rm eff}\right)^2 }  .
\end{equation}
Due to the peaked structure of $S_{\delta q}(\omega)$, one has $\omega^4 \simeq \omega_{LC}^4$ in Eq.~(\ref{speclvolt}), and therefore the calibration factor for the two voltage noise measurements is practically the same, implying that the condition for a faithful, direct, spectral measurement of the LC resonator photon occupancy reads
\begin{equation}\label{eqcondi}
  S_{imp}^{C,L} \ll \frac{q_{\rm zpf}^2}{C(x_s)^2}S_{\delta q}^{\rm peak}.
\end{equation}
We also notice that, again due to the peaked form of $S_{\delta q}(\omega)$, one has $\langle \delta \phi^2\rangle \simeq \langle \delta q^2\rangle$ [see Eqs.~(\ref{deltaq})-(\ref{deltaphi})] and therefore
\begin{equation}\label{ulc3}
  \bar{n}_{LC}^{\rm eff} \simeq \langle \delta q^2 \rangle -\frac{1}{2}.
\end{equation}

If we consider experimentally achievable parameters, enabling to approach the quantum regime for the rf circuit, $\bar{n}_{LC}^{\rm eff} \simeq 1$, one sees that the condition of Eq.~(\ref{eqcondi}) is nontrivial to satisfy, because its right hand side is of the order of $10^{-20}$ V$^2/$Hz. In fact, in this regime the charge noise spectrum peak is flattened and broadened because $\gamma_{LC}^{\rm eff}$ becomes larger and larger. Under these conditions the resonance peak height becomes comparable to the background noise level and a direct measurement of the rf photon occupancy becomes hard. Nonetheless, this charge spectrum detection method is certainly able to detect a significant laser cooling of the circuit.

One can alternatively use the spectral analysis of the system stationary state to get an indirect experimental detection of the rf circuit cooling process. In fact, one can always probe the linear response of the system by driving the rf circuit with a tunable AC voltage $V_{AC}(\omega)$, small enough in order not to modify its working point, but at the same time larger than Brownian, Johnson-Nyquist and radiation pressure noises. From Eq.~(\ref{eqq}) one has
\begin{equation}\label{eqq2}
\delta q (\omega) \simeq \chi_{LC}^{\rm eff}(\omega) V_{AC}(\omega),
\end{equation}
that is, one directly measures the effective susceptibility of the LC circuit, and in particular its FWHM $\gamma_{LC}^{\rm eff}=\gamma_{LC}+\Gamma_{LC}$ [see Eq.~(\ref{chiLCeff2})]. However, such a measurement provides also an indirect measurement of the rf photon occupancy in a large parameter regime, i.e., when the Johnson-Nyquist spectral contribution dominates over the mechanical and radiation pressure ones in the charge noise spectrum of Eq.~(\ref{spectrasusc}). In fact, in this regime, one has simply [see also Eq.~(\ref{nmeffLC})]
\begin{equation}\label{nmeffLC2}
  \bar{n}_{LC}^{\rm eff}\simeq \frac{\gamma_{LC} }{\gamma_{LC}^{\rm eff}}\bar{n}_{LC} ,
\end{equation}
that is, the temperature of the rf circuit is scaled down by the ratio $ \gamma_{LC}/\gamma_{LC}^{\rm eff}$.

An alternative way to probe the system properties is to detect the output of the optical cavity. However, any optical cavity mode interacts directly only with the mechanical resonator, and therefore it detects the dynamics of the rf circuit only indirectly, via its effects on the mechanical motion. As it is customary in cavity optomechanics \cite{RMP}, the resulting optical output spectra allows a good measurement of the effective mechanical occupancy, from which however it is hard to extract direct information about the steady state of the rf circuit.

\section{Conclusions}
\label{conc}

We have investigated a tripartite optoelectromechanical system formed by an optical cavity, a mechanical oscillator, and a MHz rf resonator which, due to its low resonance frequency, is normally in a thermally excited state even at ultra-cryogenic temperatures. We have derived the optimal conditions for achieving ground state, sympathetic, cooling of the rf resonator, modeled as an LC circuit, by means of its interaction with the mechanical resonator cooled by the laser-driven optical cavity. This requires a large optomechanical cooperativity, and an even larger electromechanical cooperativity. Under these conditions, the LC resonator can be cooled close to its quantum ground state, as confirmed by the exact numerical results in the linearized regime around the optimal working point of the circuit.
Manipulating rf resonant circuits at the quantum level would be extremely useful for the quantum-limited detection of weak rf signals, such as those employed for positioning, timing, and for the sensitive detection of rf signals of astrophysical nature.

\section*{Acknowledgments}

We acknowledge the support of the European Union Horizon 2020 Programme for Research and Innovation through the Project No. 732894 (FET Proactive HOT), of the the Project QuaSeRT funded by the QuantERA ERA-NET Cofund in Quantum Technologies, the support of the University of Camerino UNICAM through the research project FAR2018 and of the INFN through the tHEEOM-RD project. P. Piergentili acknowledges support from the European Union's Horizon 2020 Programme for Research and Innovation under grant agreement No. 722923 (Marie Curie ETN - OMT).

\appendix

\section{Pull-in voltage}

As discussed in the main text in Sec.~\ref{working}, soon after Eq.~(\ref{omegaren}), we cannot apply a too large value of the DC voltage bias $V_{\rm DC}$ due to the pull-in effect of the electrode in front of the metalized membrane, softening the intrinsic spring constant of the membrane mechanical mode. The quantity $\omega_m^2$ of Eq.~(\ref{omegaren}) must be always positive, and using Eq.~(\ref{eqxs}), one can rewrite the stability condition of Eq.~(\ref{omegaren}) as
\begin{equation}\label{omegaren2}
  \frac{m\omega_0^2(h_0\!+\!3 x_s)\!+\!\hbar n_{\rm cav}\!\left[2 \omega'(x_s)\!+\!\omega''(x_s)(h_0\!+x_s)\right]}{h_0\!+\!x_s}> 0,
\end{equation}
where $\omega'(x_s)$ and $\omega''(x_s)$ denote respectively the first and second order derivatives of the cavity frequencies with respect to x. The denominator $h_0+x_s$ is always positive because it is just the distance between the two electrodes of the effective plane-parallel capacitor modeling the membrane capacitor, so that the stability condition is equivalent to impose the positivity of the above numerator. However, it is possible to verify that the static radiation pressure frequency shift proportional to $n_{\rm cav}$ is always negligible with respect to that of electrostatic origin under typical experimental values, and therefore one gets the very simple stability condition
\begin{equation}\label{stabvol}
  x_{s} > -\frac{h_0}{3}.
\end{equation}
Using Eq.~(\ref{eqxs}) without the negligible radiation pressure term, the critical point $x_s = -h_0/3$ can be re-expressed as a condition for the maximum applicable voltage, which is given by
\begin{equation}\label{pullin}
V_{\rm pull}=\sqrt{\frac{8 m \omega_0^2 h_0^3}{27 \varepsilon_0 A_{\rm eff}}},
\end{equation}
which can be rewritten as a condition on the maximum electrical field within the membrane capacitor
\begin{equation}\label{pullmax}
\left(\frac{V_{\rm DC}}{h_0}\right)_{\rm max} =\sqrt{\frac{8 m \omega_m^2 }{27 C_m(0)}}.
\end{equation}
\\

\section{Explicit expressions in the case of a membrane-in-the-middle setup}

We have not specified in the text the explicit form of the function $\omega(x)$, which is responsible for the radiation pressure coupling between the optical mode and the mechanical resonator. In fact, the results shown in the main text can be applied to a generic geometry of the optoelectromechanical setup. However, here we provide more details for the membrane-in-the-middle case, based on the treatment of Ref. \cite{Biancofiore}. One can always express the frequency of a chosen cavity mode in the presence of a semi-transparent membrane with intensity reflectivity $R$, placed at the static position $z_0$ along the cavity axis, as
\begin{equation}\label{eq:freq-membr-shift}
   \omega(x) = \omega_c + \Theta \frac{c}{L_c}\arcsin \left\{\sqrt{R}\cos\left[2k(z_0+x)\right]\right\},
\end{equation}
where $L_c$ is the cavity length, $k = \omega_c/c$ is the wave vector associated with the chosen cavity mode, and $\Theta$ is the overlap parameter, $0\leq \Theta \leq 1$, quantifying the transverse overlap between the chosen optical and membrane vibrational modes.

The first order derivative determines the optomechanical coupling according to Eq. (\ref{optocoup}), and it is given by
\begin{equation}\label{eq:freq-membr-shift2}
   \!\!\frac{\partial \omega}{\partial x}(x)\!=\!\! -\Theta \frac{2\omega_c}{L_c}\!\sin\!\left[2k(z_0+x)\right]\!\!\sqrt{\!\frac{R}{1\!-\!R \cos^2\!\left[2k(z_0+x)\right]}}.
\end{equation}
The second order derivative instead enters into the expression for the renormalized mechanical frequency of Eq. (\ref{omegaren}) and it is given by
\begin{eqnarray}
\!\!\!\!\frac{\partial^2 \omega}{\partial x^2}(x)\!\!\!&=\!-\Theta \frac{4\omega_c^2}{c L_c}\!\!\sqrt{R}\cos\left[2k(z_0+x)\right]\frac{1-2R +R\cos^2\left[2k(z_0+x)\right]}{\left\{1-R \cos^2\left[2k(z_0+x)\right]\right\}^{3/2}}.
\end{eqnarray}\label{eq:freq-membr-shift3}

\end{document}